\documentclass[a4paper,11pt]{article}
\pdfoutput=1 

\usepackage{jinstpub} 

\usepackage{soul}

\usepackage{hyperref}

\usepackage{lineno}
\usepackage{multirow}
\usepackage{caption}
\usepackage{subcaption}
\usepackage{eqnarray,amsmath}
  
\usepackage{amssymb}

\title{A novel polystyrene-based scintillator \\ 
production process involving additive manufacturing}

\author[a,b,c,1]{S.~Berns,}
\author[d,1]{A.~Boyarintsev,}
\author[a,b,c,1]{S.~Hugon,}
\author[e,1]{U.~Kose,}
\author[e,*,1]{D.~Sgalaberna,}

\author[e]{A.~De Roeck,}
\author[d]{A.~Lebedynskiy,}
\author[d]{T.~Sibilieva,}
\author[d]{P.~Zhmurin}

\affiliation[a]{Haute Ecole Sp\'ecialis\'ee de Suisse Occidentale (HES-SO), CH-2800 Del\'emont, Route de Moutier 14, Switzerland}
\affiliation[b]{Haute Ecole d'Ing\'enierie du canton de Vaud (HEIG-VD), CH-1401 Yverdon-les-Bains, Route de Cheseaux 1, Switzerland}
\affiliation[c]{COMATEC-AddiPole, CH-1450 Sainte-Croix, Technopole de Sainte-Croix, Rue du Progr\`es 31, Switzerland}

\affiliation[d]{Institute for Scintillation Materials NAS of Ukraine (ISMA), Kharkiv 61072, Ukraine}
\affiliation[e]{European Organization for Nuclear Research (CERN), 1211 Geneva 23, Switzerland}
\affiliation[*]{Now at ETH Zurich, Institute for Particle Physics and Astrophysics, CH-8093 Zurich, Switzerland}

\emailAdd{umut.kose@cern.ch}
\emailAdd{davide.sgalaberna@cern.ch}
\emailAdd{siddartha.berns@heig-vd.ch}
\emailAdd{sylvain.hugon@heig-vd.ch}
\emailAdd{boyarintsev@isma.kharkov.ua}

\note{Corresponding author}

\abstract{
Plastic scintillator detectors are widely used in particle physics thanks to the very good particle identification, tracking capabilities and time resolution.
However, new experimental challenges and the need for enhanced  performance require the construction of detector geometries that are complicated using the current production techniques. 
In this article we propose a new production technique based on additive manufacturing that aims to 3D print polystyrene-based scintillator.
The production process and the results of the scintillation light output measurement of the 3D-printed scintillator are reported. 
}


\begin{document}
\maketitle
\flushbottom

\section{Introduction}
\label{sec:intro}

%
%

Plastic scintillator detectors were developed in the early 50s \cite{Invent-plast-scint} and are currently widely used in high-energy physics, astroparticle physics as well as in many applications like muon tomography and hadron therapy. 

Plastic scintillator is composed of a mixture of carbon and hydrogen based molecules. 
The capability of efficiently producing scintillation light by traversing charged particles makes this detectors very suitable for particle identification (PID) and allows measuring 
the particle interaction time with sub-nanosecond precision.
Plastic scintillator detectors can be used for particle tracking, see 
e.g.~\cite{t2k-fgd}. Depending on the granularity, such
detectors can reach a spatial resolution 
of better than $100 ~\mu \text{m}$ \cite{LHCb-scifi}.
Such detectors can also perform precise calorimetric measurements if alternated with layers of heavier material like iron or lead \cite{calice}.
Plastic scintillator detectors are also often used in neutrino experiments for the good tracking and PID capabilities, and with a large mass acting as active neutrino target \cite{t2k,minerva,minos}.
Thanks to the presence of low atomic mass nuclei, a unique feature of plastic scintillator detectors is the very high neutron detection efficiency, which is particularly important for neutrino experiments. Fast neutrons, with energies in the MeV range can transfer a relatively large amount of energy when scattering off hydrogen protons or when breaking-up the carbon nucleus, and allow for a neutron time of flight measurement \cite{fast-neutron-neutrino}. If the plastic scintillator is doped with, or alternated with layers composed by nuclei like lithium, boron or gadolinium, neutrons can also be thermalized and efficiently captured with a resulting release of a large number of photons \cite{solid}.

All the features of plastic scintillator-based detectors described above can be achieved often by implementing complex geometries which put important constraints on the production and detector construction processes.
In this article we describe the current state of the art of plastic scintillator production and propose a new production technique based on additive manufacturing applied to polystyerene-based plastic scintillator, with the goal to allow for an easy production but at the same time keeping performances comparable to more standard production techniques like extrusion or injection molding.

%
%

\section{Basic principles of plastic scintillator detectors}
\label{sec:plast-scint-principles}

Typically, plastic scintillator are made of polystyrene or vinyltoluene material.
Molecules of an activator, like paraterphenyl, are introduced into the polymer at a level of approximately 2\% by weight. 

This composition practically did not undergo any changes in the last fifty years, although many new polymers have been proposed.
The main hurdles are the difficulty of synthesizing big scintillator objects with a sufficient degree of transparency, the absence of a luminescent chromophore in their composition groups, the impossibility of dissolving luminescent additives to the required amount, and the large costs of synthesising the initial monomer molecules. 

\begin{figure}
\centering
\includegraphics[width=10cm]{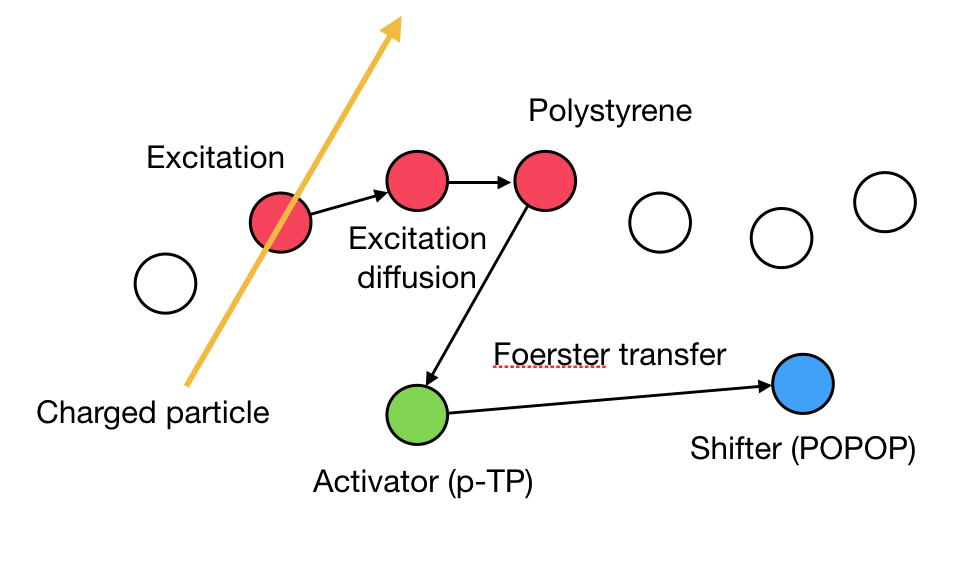}
\caption{\label{fig:scint-energy-transfer} Energy transfer in plastic scintillator. An electron excites the polystyrene molecules that hereafter diffuse the excitation to the other molecules nearby. Through the Ferster mechanism the energy is transferred to the activator molecules (e.g. p-terphenyl) that emit ultraviolet photons.
A second component (e.g. 2,2- p-phenilene-bis(5-pheniloxazole)) shifts the light wavelength from $\sim 340~\text{nm}$ to $\sim 400~\text{nm}$.}
\end{figure}

The scintillation mechanism goes through several steps.
A sufficiently energetic charged particle interacting with the plastic scintillator excites the polymer matrix molecules. The excitation energy is transferred to the activator via a resonant dipole-dipole interaction, called Foerster mechanism \cite{foerster-mechanims}, inversely proportional to the 6-th power of the intramolecule distance. This interaction strongly couples the polymer base and the activator, sharply increasing the light yield of the plastic scintillator and reducing the light emission delay. The activator must provide the necessary non-radiative collection of the polymer-base excitation energy in order to achieve a sufficient conversion into the luminescence of the required spectral range. The most important requirement for the choice of the activator molecule is the effective overlap between its absorption band and the luminescence band of the polymer to ensure an efficient transfer of excitation energy. 
The scintillation mechanism in organic materials is detailed in \cite{birk1,birk2}.
A second dopant, called 
shifter,
is usually added to change the light wavelength to a different spectrum region to provide the maximal transparency of the material to the emitted light.
Fig.~\ref{fig:scint-energy-transfer} shows schematically the mechanism via which a charged particle transfers part of its energy to the scintillator material and how the scintillation light is produced. \\

%
%

\section{Plastic scintillator detectors and three-dimensional printing}
\label{sec:plast-scint-det-3d-print}

The light produced by the plastic scintillator is usually in the blue band. 
PhotoMultipliers Tubes (PMTs) or Silicon PhotoMultipliers (SiPM) can be used to count the number of photons produced.
They can be either directly coupled with the plastic scintillator or, in more complex geometries, the light wavelength can be shifted in the wavelength shifting (WLS) fiber and shifted, for example, to the green band.
The WLS fiber can consequently capture the scintillation light and guide it to the photocounter.
The functioning of scintillating fibers, whose core is made of polystyrene-based scintillator, is very similar.


Requirements on the performance of plastic scintillator detectors include: high scintillation light yield, high transparency, fine granularity if combined with optical isolation, long-term stability and fast scintillation light production. 
Recently, very fine 3D-granular plastic scintillator detectors (3D-cube scintillator) have been proposed for future neutrino detectors. One example is given by an active neutrino target detector, made of two million $10\times10\times10~\text{mm}^3$ plastic-scintillator cubes, optically isolated and with three independent holes along the three orthogonal directions that host WLS fibers \cite{superfgd}. 
Future step changes in this technology may require a larger target mass combined with finer granularity.
An obvious limitation is the challenge to assemble so many small cubes with the required precision.

\subsection{Production processes}
\label{sec:plast-scint-prod-std}


High-quality plastic scintillator can be produced with the cast polymerization technique \cite{cast-polymerization}. A liquid monomer with dissolved dopants is poured into a mold and heated. Eventually, after cooling, a rigid solid plastic is obtained. This technique is quite expensive compared to other methods, given the rather complex technique and the relatively long production time required. 

Another possible technique is the injection molding \cite{injection-molding-1,injection-molding-2}.  This method is widely used in industry: optically transparent granulated polystyrene is mixed with the dopant and directed into the mold at a temperature of approximately $200^\circ$C.  The quality of the plastic scintillator is inferior to the one obtained with cast polymerization, but sufficient to make particles detectors for most of the applications. 

An alternative technique is the extrusion \cite{extrusion-1,extrusion-2}. This production method can provide plastic scintillator at very low cost but with usually poorer optical attenuation properties. It is often used for the production of very large detectors. \\
The methods described above allow to produce individual plastic scintillator components that afterward need to be assembled. However, when more complex geometries are needed, these methods may be insufficient or the assembly may become very difficult. 
An ideal solution would consist of a production technique that fabricates all the individual plastic scintillator objects of the desired gemeotrical shape already assembled, composing a single scintillator block, and at
the same time keeping all the quality characteristics high as
described above. 

%
%

\subsection{Additive manufacturing}


Plastic scintillator detectors are traditionally produced through the techniques described in sec.~\ref{sec:plast-scint-prod-std}, and then manufactured and adapted to the detector geometry by a subtractive process such as machining or drilling. With such process it is possible to manufacture single cubes or internal cells with excellent accuracy and good transparency but, depending on the final detector geometry, with difficulties in the assembly or machining.
Additive manufacturing (AM) opens a door to new automated processes that can drastically simplify the construction of plastic scintillator detectors. 

More commonly known as rapid prototyping or three-dimensional (3D) printing, AM processes are used to design and create parts by adding material in a layer-wise fashion. 
AM has undergone significant developments since its conception for the first commercial application for the 
technology, known as stereolithography \cite{am-invention}. 
%
%
AM technologies are capable of fabricating parts for mass customization with multiple materials and complex internal geometries. Moreover, these processes allow cheaper and quicker operations compared to conventional techniques because of a tool-less operation.

\begin{figure}
\centering
\includegraphics[width=14cm]{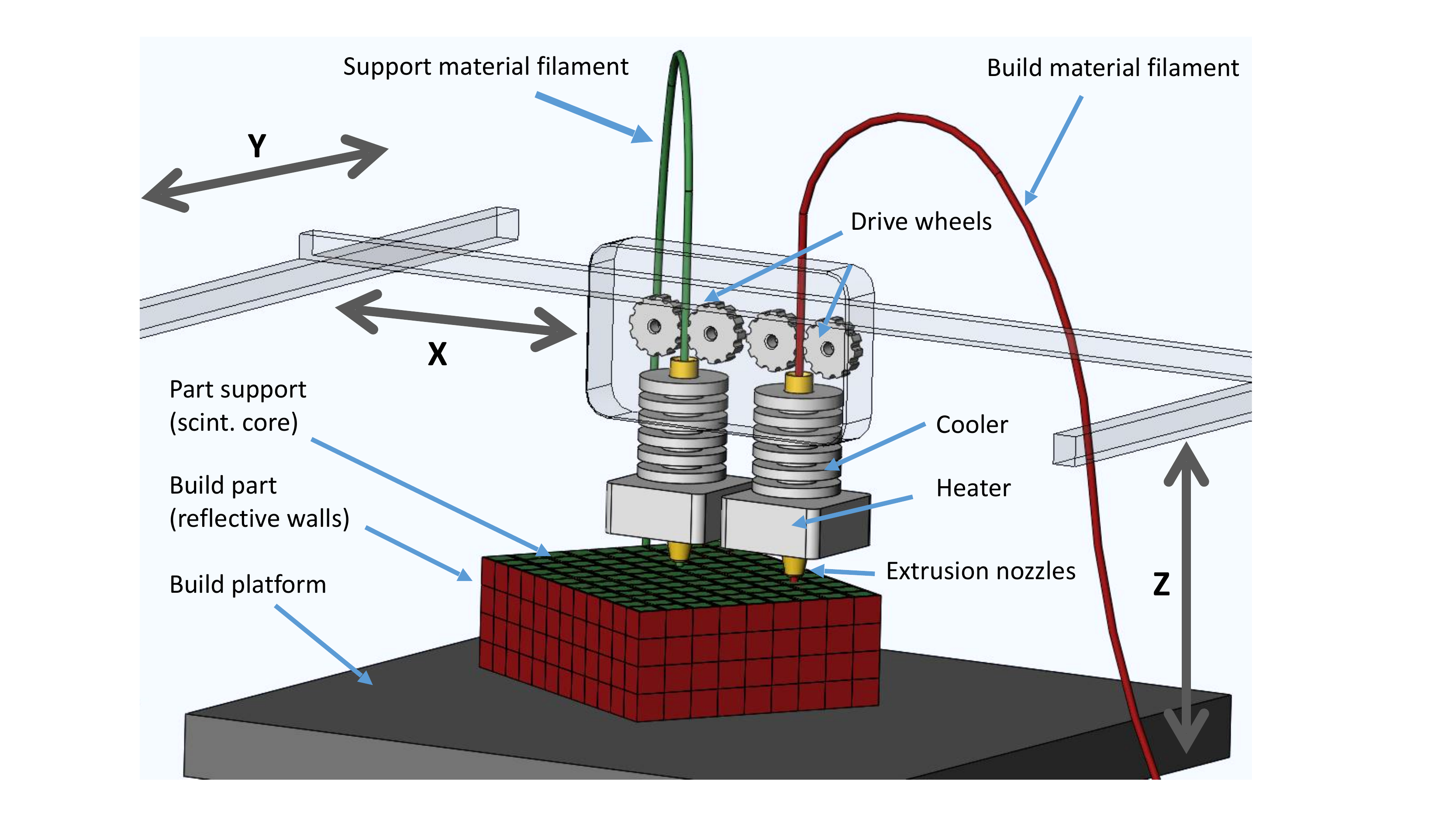}
\caption{\label{fig:fdm} Representation of the FDM process. 
}
\end{figure}

The Fused Deposition Modeling (FDM) process \cite{fdm} is one of the most common AM processes.
It consists of material deposition line by line, building many layers until the target volume is achieved. 
A thermoplastic filament in the form of thin wires is used as the base material which passes through a set of feeding rollers into the liquefier \cite{fdm-description,astm-am-definition}.
The plastic is heated to its melting temperature with the help of heaters and squeezed out of the nozzle tip, with typical diameter between 0.2 and 0.8 mm. 
This process can produce complex parts with the help of support structures. Multi-material fabrication, to combine different materials in different orientations, is also possible.
The advantages of this method are: fast production, ability to print transparent materials, simultaneous multi-material printing, and ability to treat standard polymers and low cost.
The basic principle of the FDM process is shown in fig.~\ref{fig:fdm}. 
FDM has a strong potential for successfully producing plastic scintillator detectors with complex geometries.

In summary, if additive manufacturing of multiple materials can be successfully used for plastic scintillator, it will allow  to produce detectors similar to the 3D-cube scintillator detector described in sec.~\ref{sec:plast-scint-det-3d-print} 
easily as a single block of material, where the pattern of active scintillator volumes covered by an optical reflector is 
handled directly by the 3D printer.
Since current 3D printers can definitely make object bigger than $20\times20\times20~\text{cm}^3$, more than 8,000 cubes of $1~\text{cm}^3$ each could be manufactured at the same time.

%
%

\section{3D-printing of polystyrene-based scintillator}


Initial trials for 3D printing of scintillators were done  using the stereolithographic technique \cite{3dprinted-scint-first} or fused deposition modeling \cite{3dprinted-scint-first-fdm}. Although these showed for the first time the strong potential of this approach, the quality was not good enough for applications such    as for particle detectors. 

Our goal is to apply the FDM process to 3D print polystyrene scintillator objects.
This strategy profits from the fact that   FDM is a well established technique that can easily 3D print multimaterials, mandatory for tracking detectors with components that need optical isolation.
Moreover, polystyrene is a very common scintillator material used in most of the high-energy physics experiments, with perfect quality characteristics  for particle detectors, like a large amount of scintillation light yield, excellent transparency, a fast scintillation process and last but not least, a low cost.

The first test showing the potential of the FDM process 
was done by simply melting  pre-machined polystyrene scintillator. The obtained good transparency indicated that FDM could be used with polystyrene scintillator. Indeed it was 
known that polystyrene can be 3D printed, however it was not clear whether sufficient transparency could be obtained.
It was also found that  temperatures of around $200^\circ$C allowed to melt the scintillator without degradation of the material properties.
Reducing impurities or the presence of air to avoid oxidation helps to improve the performance.

The preparation for FDM includes the production of the filament and the optimisation of the FDM printer parameters.
At the end, during this study, different polystyrene scintillator cubes were produced and tested.

%
%

\subsection{Filament production}
\label{sec:filament}

\begin{figure}
\centering
\raisebox{-0.5\height}{\includegraphics[width=8cm]{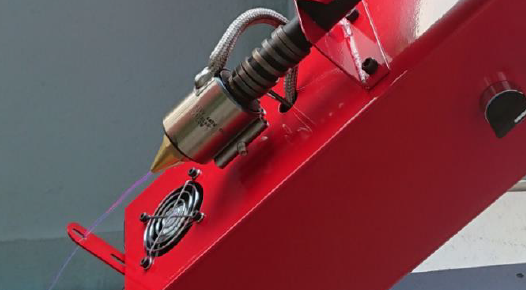}}
\caption{\label{fig:extruder} Picture of the extruder used to produce the scintillator filament.}
\end{figure}

The plastic scintillator filament is made of polystyrene doped with 2\% by weight of p-terphenyl (PTP) and 0.05\% by weight of 2,2- p-phenilene-bis(5-pheniloxazole) (POPOP).
It was obtained by using an extruder, shown in fig.~\ref{fig:extruder}.
The main challenge in the FDM filament production is  the hardness of polystyrene, causing the 3D-printer rollers to crush the filament when inserted, or to produce cracks in the filament when positioning the 3D-printer extruder before starting the process. Moreover, a pure polystyrene filament cannot be rolled on a standard spool.

To overcome this problem, different solutions were tested.
We tried to produce the filament from a copolymer of styrene and butadiene, i.e. styrene-butadiene-styrene thermoplastic elastomers (SBS TPE plastic). 
The butadiene makes the plastic more flexible. 
However, POPOP and PTP are poorly soluble in this plastic and, as a consequence, the scintillator light yield is reduced down to 40\% of the standard polystyrene scintillator. 

A second solution consists of adding a plasticizer component to the polystyrene, to increase the mobility of structural elements in the polymer by reducing the crystallizability, the van der Waals surface energy (intermolecular forces between polymer chains), the glass transition temperature 
and rigidity. 
It also increases the toughness, i.e. the ability of a material to absorb energy and plastically deforming without creating fractures.
After the copolymerization, a flexible filament can be obtained.

\begin{figure}
\centering
\includegraphics[width=10cm]{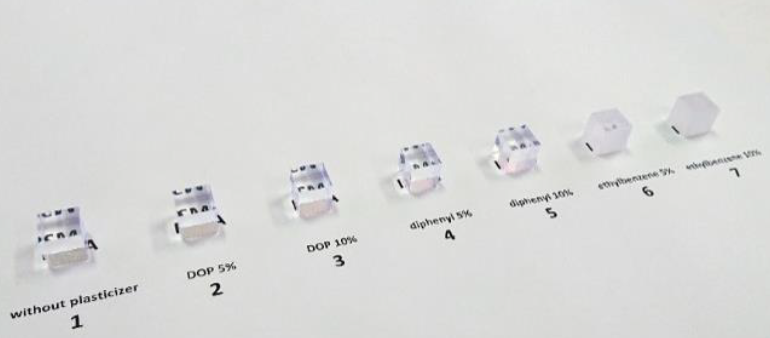}
\caption{\label{fig:plasticizer} Cubes of cast plastic scintillator with an addition of different plasticizer, in the following order from left to right: without plasticizer, 5\% DOP, 10\% DOP, 5\% byphenil, 10\% byphenil, 5\% ethylbenzene, 10\% ethylbenzene. The fractional amount of plasticizers is by weight. }
\end{figure}

The selection of the plasticizer was made according to: flexibility, transparency, light yield and long-life of the plastic scintillator.
Drawbacks could potentially be the loss of transparency, the aging and the loss of light output.
The plastic samples were produced with the cast technology and different types of plasticizer were tested. 
Fig.~\ref{fig:plasticizer} shows the different types of plasticizer that were tested.
The Compton edge of the spectra obtained by exposing different scintillator samples to a $^{137}\text{Cs}$ $\gamma$-source was measured and used to compare the relative light yield. 

\begin{figure}
\centering
\includegraphics[width=10cm]{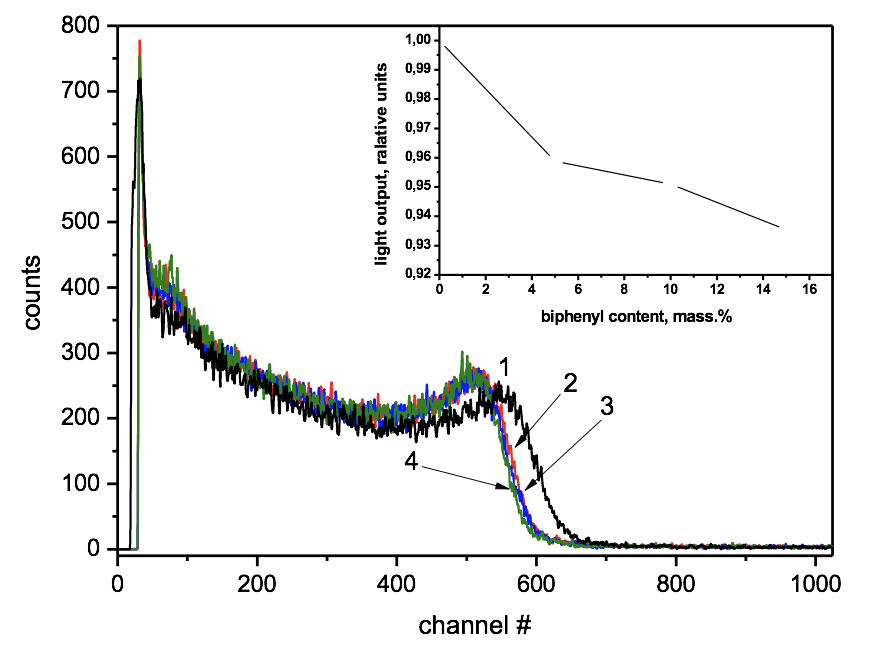}
\caption{\label{fig:byphenil-light-yield} The pulse height spectra of polystyrene-based plastic scintillator exposed to a $^{137}\text{Cs}$ $\gamma$-source is shown. 
The X axis shows the number of ADC channels.
Different byphenil compositions by weight were tested: standard composition plastic scintillator (1 - black), 5\% biphenyl (2 - red), 10\% biphenyl (3 - blue), 15\% biphenyl (4 - green). 
The inset plot on top right shows the dependence of the relative light output as a function of the byphenil content.}
\end{figure}

It was found that adding about 5\% by weight of byphenil can achieve the  goals, with a light output better than 95\% compared to pure scintillator. 
In fig.~\ref{fig:byphenil-light-yield} the light yield measured from scintillator samples with different byphenil contents is shown.
The concentration of biphenyl has no particular impact on the plastic scintillator transparency. 
According to data, the obtained mechanical properties of scintillator filament produced by hot extrusion do not considerably depend on the biphenyl content in the polystyrene matrix.
Other plasticizer, like DOP and ethylbenzene, were tested. However, it was found that already 5\% by weight of the first reduces the scintillation light yield by approximately 30\%, although providing good transparency, whilst the latter drastically increase the opacity of the scintillator.

\begin{figure}
 \centering
   \begin{subfigure}{.45\textwidth}
    \centering
   \includegraphics[width=6.7cm, height=5cm]{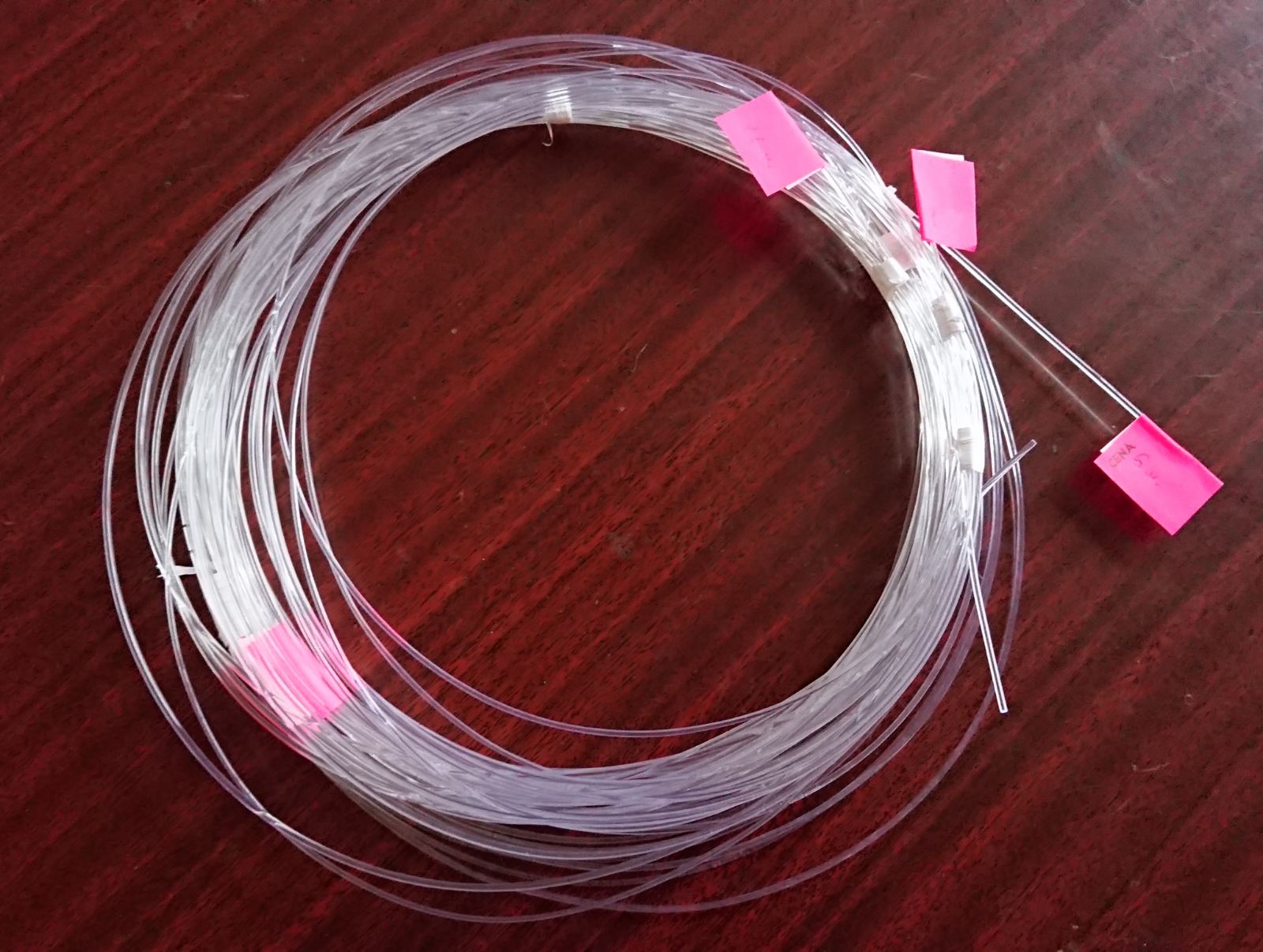} 
    \end{subfigure}
    \hspace{-0.7cm}
   \begin{subfigure}{.45\textwidth}
    \centering
   \includegraphics[width=3.5cm,height=5cm]{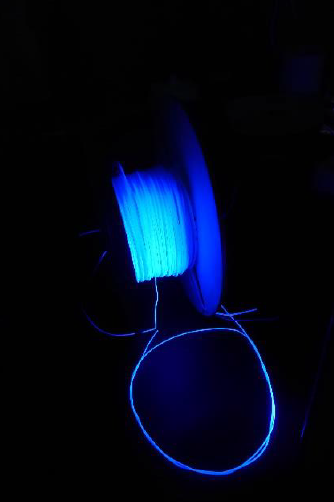} 
    \end{subfigure}
\caption{\label{fig:filament} Photographs of the polystyrene-based scintillator filament used for 3D-printing with the FDM technique. On the right picture, the filament was illuminated with UV light.}
\end{figure}

Finally, a filament with a diameter of 1.75~mm made of polystyrene-based plastic scintillator with an addition of 5\% by weight of byphenil was produced and used for 3D-printing. 
Photographs of the polystyrene scintillator filament are shown in fig.~\ref{fig:filament}.


%
%

\subsection{3D-printing process}
\label{sec:3d-printing-process}


Different samples of plastic scintillator were produced with both ``Roboze One+400'' \cite{printer-roboze} and ``CreatBot Dx2'' \cite{printer-creatbot} printers.
Overall both printers provided satisfying results, although the cubes used for the final tests were produced with the 
``CreatBot Dx2''. 
The polystyrene scintillator filament was produced following the method described in sec.~\ref{sec:filament}.

\begin{figure}
\centering
\raisebox{-0.5\height}{\includegraphics[width=6cm]{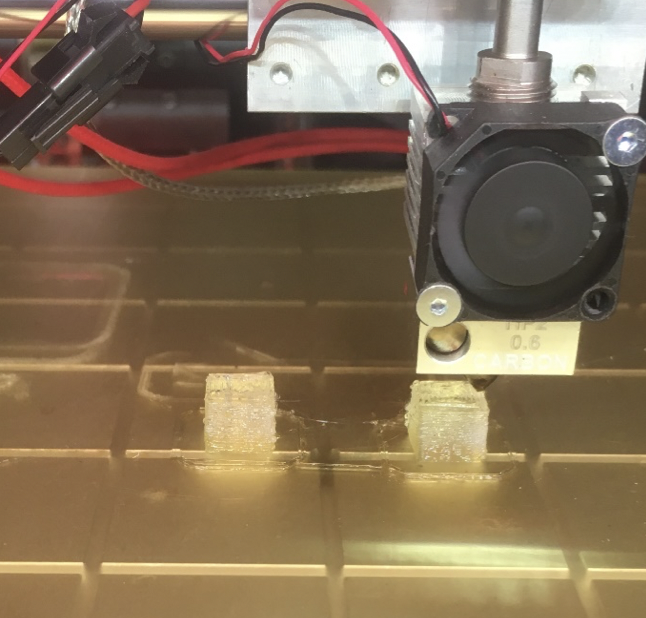}}
\caption{\label{fig:printing-scint} 3D-printing of the plastic scintillator cubes. 
}
\end{figure}

Important tuning parameters are the ratio between the extruder diameters and the layer thickness, the line overlap, the printing speed and the extruder flow. After a campaign of tuning all of those parameters, a fully dense and transparent scintillator cube was produced.
The FDM extruder diameter was set to the range between 0.4 mm and 0.6 mm. 
Extrusion temperatures between $240^{\circ}\text{C}$ and $250^{\circ}\text{C}$ and printer bed temperatures around $100^{\circ}\text{C}$ were tested. 
The printing speed was between 10~mm/s to 30~mm/s, with a layer thickness between 0.1 to 0.2 mm.
In fig.~\ref{fig:printing-scint} two polystyrene scintillator cubes are shown during the 3D-printing process. 


%

In the tests the priority was given to the optimisation of the scintillator transparency.
It was found that increasing the temperature of chamber, bed and extruder improves the scintillator transparency. On the other hand, the cubic shape is better preserved when not too high temperatures are used. 
Although improvements of the geometrical shape have not 
been optimised yet, the cubes could be produced with an edge tolerance of about 0.5~mm.
Possible further studies for future tests are discussed in sec.~\ref{sec:conclusions}.



Challenges are possible presence of small air bubbles or not fully-melted zones, that can be  a potential source of scintillation light scattering and diffusion inside the scintillator cube. In fact, this could worsen the scintillation light output uniformity within the cube or introduce some opacity. 
We found that increasing the printer extrusion temperature can minimise these features: the viscosity of the material drops down, the material flows better to the previous printed line with capillary and gravitational forces, the gaps are fully filled and air bubble are minimised.



While the inner core of the scintillator cube typically showed a very good transparency, it was noted that the outer surface was more opaque. 
A particular feature of the FDM technique is the quite high roughness of the outer surfaces vertical with respect to the 3D printer bed. This is due to the visible pattern of the melted filament that makes the outer part of the cube less transparent.
Reducing the layer thickness would squeeze the melted polystyrene and improve the transparency.
Although more studies will be performed, this feature will not be a particular issue for detectors similar to 3D-cube scintillator where the active volume has to be covered by an opaque or reflecting surface. Moreover, this effect is found only on the outermost surface of the printed scintillator, preventing the inner part (either scintillator or reflector) from being affected by this feature.



\begin{figure}
\centering
\raisebox{-0.5\height}{\includegraphics[width=6cm]{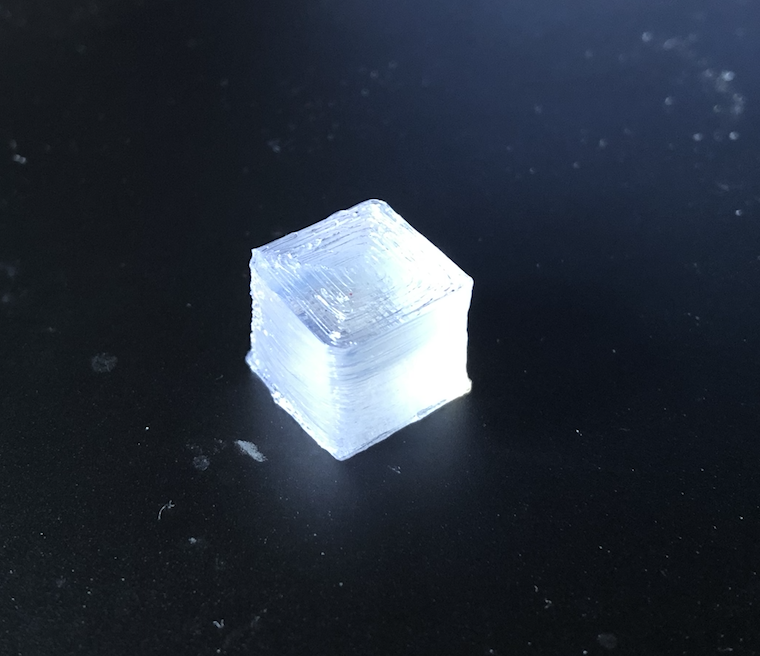}} 
\hspace{0.5cm}
\raisebox{-0.5\height}{\includegraphics[width=5.7cm]{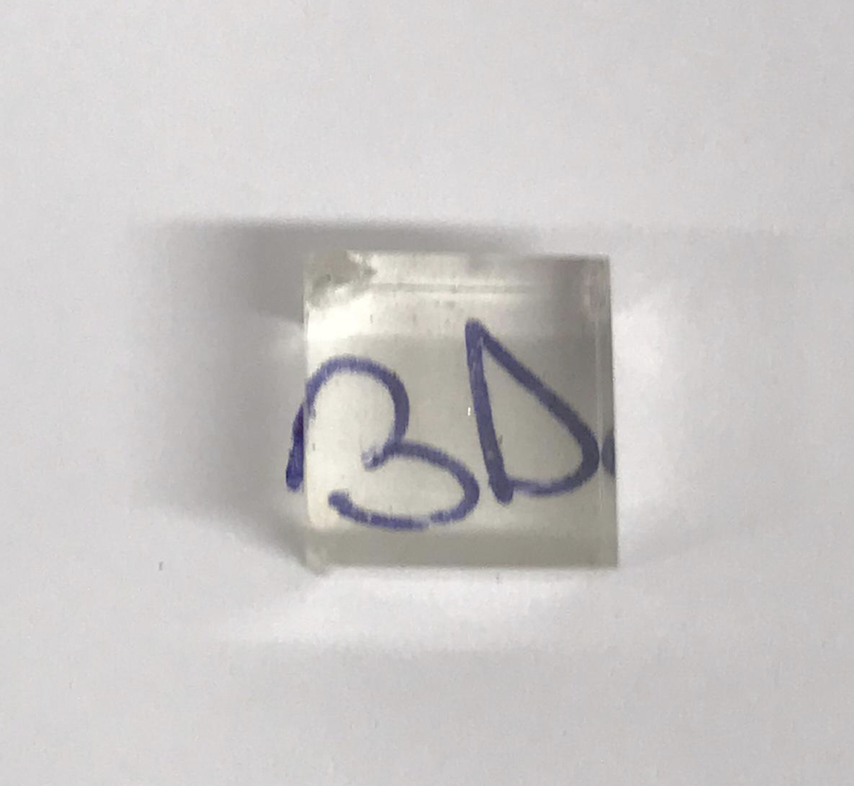}}
\caption{\label{fig:printed-cube} 3D-printed plastic scintillator cube before (left) and after (right) polishing its outer surface. 
}
\end{figure}

In order to measure the scintillator light output, the outer surface was polished leading to  a slightly smaller cube.
In fig.~\ref{fig:printed-cube} the 3D printed scintillator cube is shown before and after the polishing.

%
%

\section{Performance of the 3D printed scintillator}
\label{sec:measurements}

The performance of the 3D printer polystyrene scintillator was evaluated by measuring the scintillation light yield and counting the number of photons produced per amount of energy deposited by charged particles.
The scintillator cube was 3D printed with ``CreatBot Dx2''.
The extrusion diameter was 0.4 mm, the extrusion and bed temperatures respectively $245^{\circ}$C and $100^{\circ}$C. The layer thickness was 0.1 mm, while the 3D printing speed was 13 mm/s.
After polishing the outer surface the cube has a size of approximately $9\times9\times9~\text{mm}^3$, as described in sec.~\ref{sec:plast-scint-prod-std}. 

\begin{figure}
\centering
\includegraphics[width=6cm]{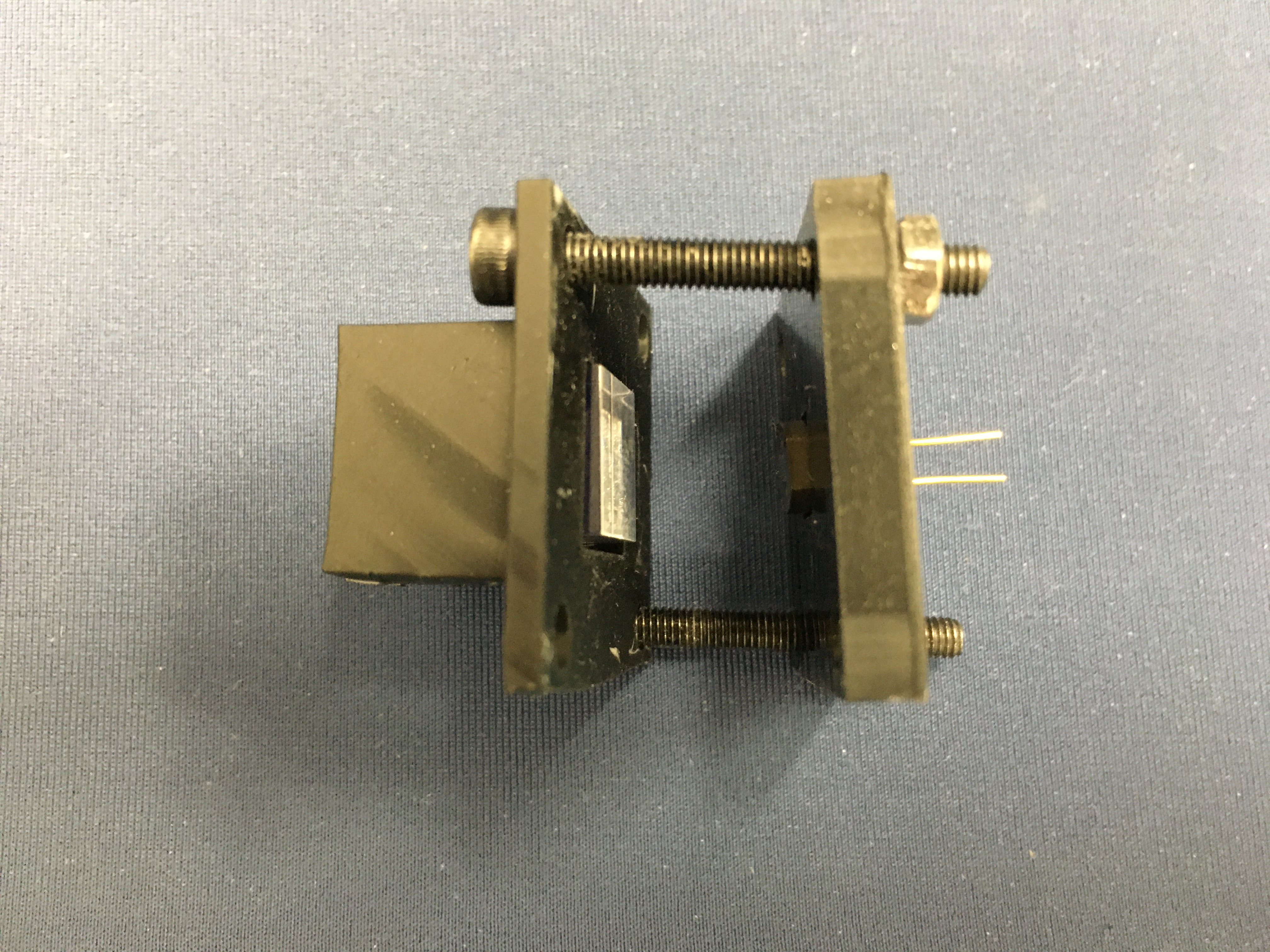}
\caption{\label{fig:test-connector} The black optical connector used to couple the MPPC with the scintillator cube is shown. The connector host the cube in a cubical hole. A 5~mm thin piece of EPDM foam is placed inside the connector hole to push the cube toward the MPPC active surface and improve the coupling. The connector was made with a stereolitography 3D printer. 
}
\end{figure}

The scintillation light yield was measured with Hamamatsu S13360-1350CS Multi-Pixel Photon Counter (MPPC). The number of photoelectrons (p.e.) was extracted after measuring the MPPC gain, defined as the number of ADC counts to the number of p.e. conversion. The MPPC active region ($1.3\times1.3~\text{mm}^2$) was directly coupled with the polystyrene scintillator cube with a black optical connector, as shown in fig.~\ref{fig:test-connector}. A piece of soft black EPDM foam was placed inside the connector to push the cube against the MPPC, improving the coupling.
The MPPC charge signal was read-out with a CAEN DT5702 front-end board \cite{caen-feb}.
This setup allows to measure mostly the direct light produced by the scintillator, whilst limiting the light reflection on the outer surfaces because the inner surface of the optical connector as well as the EPDM foam are black. 
Hence, the impact of the light attenuation is mitigated.

The scintillator cube was exposed to a strontium-90 source ($^{90}\text{Sr}$).
It emits electrons with energies up to 0.5~MeV through inverse-$\beta$ decay process, producing a yttrium-90 isotope ($^{90}\text{Y}$) that would further decay to zirconium-90 ($^{90}\text{Zr}$), producing an electron and an antineutrino in the final state. Being a three-body decay the electron can take energies mostly between 0.5~MeV and 2.28~MeV \cite{sr90-source-1,sr90-source-2}. 
The source was placed on top of the optical connector, opposite of the side with respect to the MPPC position (left side of the picture in fig.~\ref{fig:test-connector}).
Given the optical connector thickness of 2~mm, made of plastic material, mostly electrons from $^{90}\text{Y}$ and with energies above 1~MeV are expected to be able to deposit light in the scintillator cube.

\begin{figure}
\centering
\includegraphics[width=8cm]{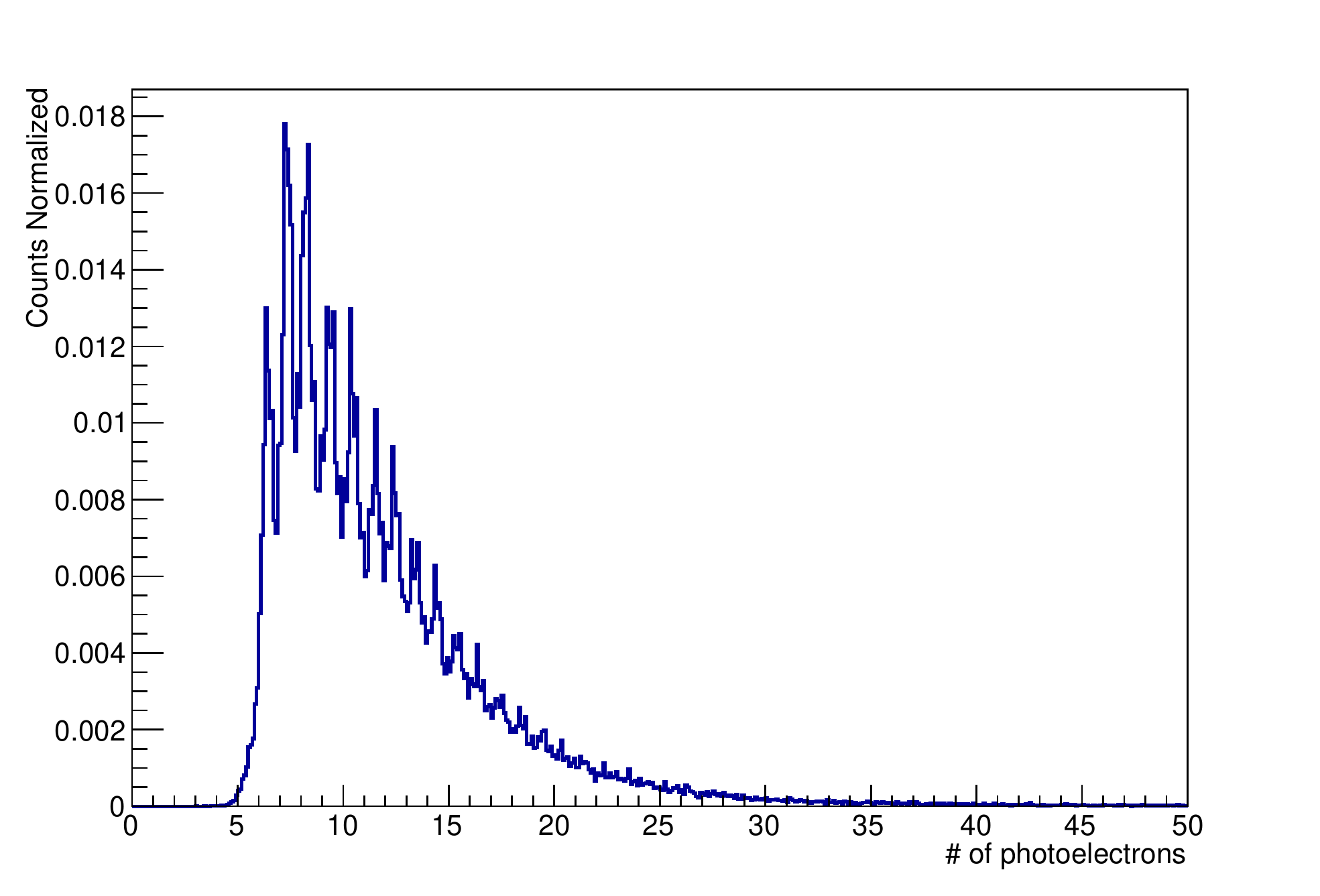}
\caption{\label{fig:test-source-3d-print-result} Normalized energy spectrum measured by exposing the 3D printed polystyrene scintillator cube to a $^{90}\text{Sr}$ source. Counts as a function of the number of MPPC p.e. is shown. The peaks correspond to different numbers of p.e. The mean number of p.e. is about 12.
}
\end{figure}

The light output performance was compared with that of plastic scintillator cubes produced with standard techniques, such as extrusion and cast (see sec.~\ref{sec:plast-scint-prod-std} for a detailed description), by the ``Institute for Scintillation Materials of the National Academy of Science of Ukraine'' (ISMA). The cast and extruded plastic scintillator samples consisted of about $10\times10\times10~\text{mm}^3$ cubes. It is worth mentioning that the scintillator composition of the cubes produced with different techniques was the same. 
The measured light yield of the 3D printed scintillator cube is shown in fig.~\ref{fig:test-source-3d-print-result}. A mean of approximately 12 p.e. was measured.
The same light yield was measured with the scintillator cubes produced using the cast and extrusion techniques, and both show energy spectra very similar to the one in  fig.~\ref{fig:test-source-3d-print-result}. 


\begin{figure}
\centering
\includegraphics[width=8cm]{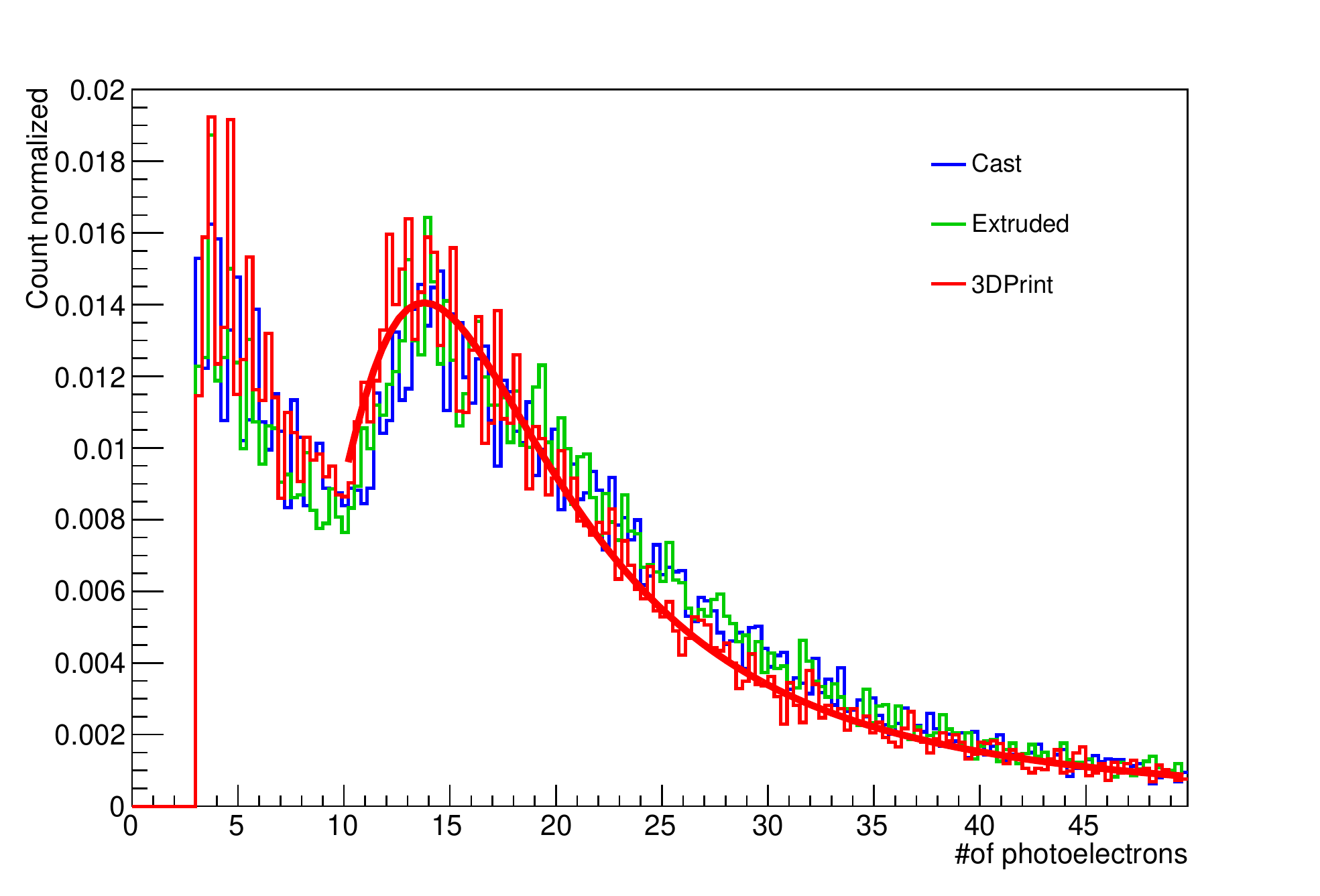}
\caption{\label{fig:test-cosmic-3d-print-result} Spectrum of cosmic muons obtained with cast (blue), extruded (green) and 3D printed (red) cubes. For illustration purpose the 3D printed cube data is fitted with a Landau function (red curve).
A cut of 3 p.e. was applied to get rid of electronic noise and dark count rate from the spectrum. 
}
\end{figure}

An independent measurement of the light yield was performed by collecting cosmic data to confirm the results obtained with the $^{90}\text{Sr}$ source.
Cosmic particles deposit about 2 MeV/cm in plastic scintillator.  
The same setup as for the $^{90}\text{Sr}$ source test was used.
In fig.~\ref{fig:test-cosmic-3d-print-result} the energy spectra, obtained by the 3D-printed, extruded and cast polystyrene-based scintillator samples, are compared.
The spectra were fitted with a Landau function, excluding the low-energy region largely contaminated by environmental photons. 
It was found that the 3D-printed scintillator shows a light yield similar to the one obtained with the extruded and cast samples, confirming the results of the $^{90}\text{Sr}$ source test. 
The most probable and width values of the fitted Landau distribution of the 3D-printed cube are slightly smaller than the ones from extruded and cast samples, as shown in table~\ref{tab:fit-parameters-cosmic}. 
Additional uncertainties on the measured parameters could be introduced by environmental photons contaminating the region dominated by cosmic particles.
These results are consistent with the $^{90}\text{Sr}$ source ones within the systematic uncertainties of the measurement.

\begin{table}[h!]
\caption{\label{tab:fit-parameters-cosmic} Parameters obtained by fitting the cosmic data in fig.~\ref{fig:test-cosmic-3d-print-result} with a Landau function.}
\begin{center}
\begin{tabular}{ |c|c|c|c| } 
\hline
 & Most Probable Value (p.e.) & Width (p.e.) \\
  \hline
  Cast      & 15.5  & 3.7 \\ 
  Extruded  & 15.4  & 3.7 \\ 
  3D print  & 14.5  & 3.3\\ 
\hline
\end{tabular}
\end{center}
\end{table}

Although the measured number of photoelectrons can be affected by differences in the experimental setup (e.g. cubes size, potentially different quality of the cube surface polishing or small differences in the coupling between the cube and the MPPC, etc.) and that a limited number of samples has been tested, we can assert that we were able to 3D print polystyrene scintillator of relatively good quality with light yield performance similar to the ones available on the market and widely used (and required) for particle detectors. 
It is worth mentioning that this experimental setup does not aim to evaluate the effect of the attenuation length in scintillator to the overall light output; this will be measured in  future tests. \\

%
%

\section{Conclusions}
\label{sec:conclusions}


In this article we proved the possibility of 3D printing a polystyrene-based scintillator with fused deposition modeling.
This process requires both materials and production techniques that are commercially available and is potentially optimal for constructing detectors made of complicated geometries.
The scintillation performances of this material are very well known and described in literature.

We developed the entire process, including the production of the filament made of polystyrene scintillator and the tuning of the 3D printing parameters. 
We found that melting polystyrene scintillator does not significantly deteriorate neither the transparency nor the light yield.
The scintillation light yield was found to be comparable to the one given by cast or extruded scintillator, based on a limited number of samples that were tested.
A detailed characterisation of the 3D printed scintillator will be performed in the future with a large number of samples to test the reproducibility and precisely measure the scintillator properties, such as attenuation length, time decay constant and aging effects.

In order to be able to 3D print a particle detector made of 3D-cube scintillator, 
as described in sec.~\ref{sec:plast-scint-det-3d-print}, 
further R\&D steps are needed.
First, the scintillator attenuation length will be studied, with attempts of improving it whilst maintaining good geometrical tolerances.
Optimisations will be made on the tuning of operational parameters such as the printer bed, room and extruder temperatures.
A further step will be the simultaneous 3D printing of scintillator and optical reflector. 
Finally, the cube will have to be made with holes along the three orthogonal directions to host the WLS fibers.

The results reported in this article are a first milestone of an ongoing R\&D program which aims to adopt an additive manufacturing techniques for the production of scintillator-based particle detectors.







\acknowledgments
We thank Gianmaria Collazuol for fruitful discussions.




\begin{thebibliography}{99}

\bibitem{Invent-plast-scint} M. G. Schorr and F. L. Torney, Phys Rev, 1950, 80, 474-474

\bibitem{t2k-fgd} P.A. Amaudruz et al., ``The T2K Fine-Grained Detectors'', Nucl.Instrum.Meth. A696 (2012) 1-31

\bibitem{LHCb-scifi} C. Joram et al., ``LHCb Scintillating Fibre Tracker Engineering Design Review Report: Fibres, Mats and Module'', CERN-LHCb-PUB-2015-008, 03/2015

\bibitem{calice} V. Andreev et al., ``A high-granularity plastic scintillator tile hadronic calorimeter with APD readout for a linear collider detector'', Nucl.Instrum.Meth.A564:144-154,2006

\bibitem{t2k} K.Abe et al., ``The T2K experiment'', Nucl. Instrum. Meth. A 659, 106 (2011)

\bibitem{minerva} L. Aliaga et al., ``Design, Calibration and Performance of the MINERvA Detector'', Nucl. Inst. and Meth. A743 (2014) 130

\bibitem{minos} D.G. Michael et al., "The magnetized steel and scintillator calorimeters of the MINOS experiment," Fermilab-Pub-08-126, Nucl.Instrum.Meth.A596:190-228(2008), Issue 2, 1 November 2008, arXiv:0805.3170



\bibitem{fast-neutron-neutrino} L. Munteanu et al., ``A new method for an improved anti-neutrino energy reconstruction with charged-current interactions in next-generation detectors'', arXiv:1912.01511 [physics.ins-det]

\bibitem{solid} 
Y. Abreu et al. (SoLID Collaboration), ``Performance of a full scale prototype detector at the BR2 reactor for the SoLid experiment'', 2018 JINST 13 P05005


\bibitem{foerster-mechanims} T. Foerster, Ann. Phys. 2, 55 (1948)

\bibitem{birk1} J. B. Birks, Nuclear Science, IRE Transactions on, 1960, 7, 2-11.

\bibitem{birk2} J. B. Birks, ``The theory and practice of scintillation counting'', Pergamon Press; [distributed in the Western Hemisphere by Macmillan, New York], Oxford, New York, 1964.

\bibitem{superfgd} A. Blondel et al., JINST 13, P02006 (2018), 709, arXiv:1707.01785 [physics.ins-det].

\bibitem{cast-polymerization} C.A. Harper and E.M. Petrie, ``In Plastics Materials and Processes: A Concise Encyclopedia'' (2003), John Wiley \& Sons

\bibitem{injection-molding-1} G.S. Atoian, et. al., Nucl. Instr. and Meth. A 531, 467 (2004)

\bibitem{injection-molding-2} R. Appel et. al., Nucl. Instr. and Meth. A 479, 349 (2002) 

\bibitem{extrusion-1} J.C. Thevenin, L. Allemand, E. Locci, P. Micolon, S. Palanque and M. Spiro, ``Extruded polystyrene, a new scintillator'', Nucl. Instrum. Meth. A 169 (1980) 53

\bibitem{extrusion-2} Anna Pla-Dalmau, Alan D. Bross and Kerry L. Mellott, ``Low-cost extruded plastic scintillator'', Nucl. Instrum. Meth. A  466 (2001) 482

\bibitem{am-invention} S.T. Newman, Z. Zhu, V. Dhokia, and A. Shokrani, ``Process planning for additive and subtractive manufacturing technologies,'' CIRP Ann. - Manuf. Technol., vol. 64, no. 1, pp. 467-470, 2015.


\bibitem{fdm} ``FDM Technology, About Fused Deposition Modeling'' (copyright by Stratasys) http://www.stratasys.com/3d-printers/technologies/fdm-technology. [Accessed: 26-Mar-2017].

\bibitem{astm-am-definition} M. P. Groover, Fundamentals of modern manufacturing: materials, processes, and systems. John Wiley \& Sons, Inc, 2013

\bibitem{fdm-description} A. Bellini and S. G$\ddot{\text{u}}$\c{c}eri, ``Mechanical characterization of parts fabricated using fused deposition modeling'', Rapid Prototyp. J., vol. 9, no. 4, pp. 252-264, 2003.

\bibitem{3dprinted-scint-first} Mishnayot, Y. and Layani, M. and Cooperstein, I. and Magdassi, S. and Ron, G., ``Three-dimensional printing of scintillating materials'', Rev.Sci.Instrum. 85 (2014) 085102, ArXiv:1406.4817 

\bibitem{3dprinted-scint-first-fdm} 
M. Hamel and G. Lebouteiller, ``Attempting to aprepare a plastic scintillator from a biobased polymer'', J. Appl. Polym. Sci. 2020, 137, 48724.


\bibitem{printer-roboze} \url{https://www.roboze.com/en/3d-printers/roboze-one-400.html}

\bibitem{printer-creatbot} \url{https://www.creatbot.fr/}



\bibitem{sr90-source-1} Dixon, J., Rajan, A., Bohlemann, S. et al. ``Evaluation of a Silicon $^{90}\text{Sr}$ Betavoltaic Power Source'' Sci Rep 6, 38182 (2016)

\bibitem{sr90-source-2}  M. Sadeghia, S. Yektab, H. Ghaedic and E. Babanezhadb, ``$\text{MnO}_2$ NPs-AgX zeolite composite as adsorbent for removal of strontium-90 ($^{90}\text{Sr}$) from water samples: Kinetics and thermodynamic reactions study'', Materials Chemistry and Physics 197 (2017) 113-122

\bibitem{caen-feb} \url{https://www.caen.it/products/dt5702/}







\end{thebibliography}
\end{document}